\documentclass[letterpaper,twocolumn,10pt]{article}
\usepackage{usenix2019_v3}

\usepackage{verbatim}
\usepackage{graphicx}
\usepackage[hyphens]{url}
\usepackage{hyperref}
\usepackage[htt]{hyphenat}
\usepackage{xspace}
\usepackage{booktabs}
\usepackage{forest}
\usepackage{subcaption}
\usepackage{graphicx}
\usepackage{amsmath}
\usepackage{amssymb}\usepackage{pifont}\usepackage{authblk}

\newcommand*\justify{\fontdimen2\font=0.4em\fontdimen3\font=0.2em\fontdimen4\font=0.1em\fontdimen7\font=0.1em\hyphenchar\font=`\-}

\renewcommand{\texttt}[1]{\begingroup
	\ttfamily
	\begingroup\lccode`~=`/\lowercase{\endgroup\def~}{/\discretionary{}{}{}}\begingroup\lccode`~=`[\lowercase{\endgroup\def~}{[\discretionary{}{}{}}\begingroup\lccode`~=`.\lowercase{\endgroup\def~}{.\discretionary{}{}{}}\catcode`/=\active\catcode`[=\active\catcode`.=\active
	\justify\scantokens{#1\noexpand}\endgroup
}

\usepackage{paralist}
\usepackage{enumitem}
\newcounter{descriptcount}
\newlist{enumdescript}{description}{2}
\setlist[enumdescript,1]{before={\setcounter{descriptcount}{0}\renewcommand*\thedescriptcount{\arabic{descriptcount}.}}
	,font=\bfseries\stepcounter{descriptcount}\thedescriptcount~
}

\usepackage{listings}
\usepackage{color}
\definecolor{light-gray}{gray}{0.95}
\usepackage{textcomp}

\usepackage[flushleft]{threeparttable}
\usepackage{adjustbox}
\usepackage{wasysym}
\usepackage{multirow}

\begin{document}

\date{}

\newcommand{\attack}{Flushgeist}

\title{\Large \bf  \attack{}: Cache Leaks from Beyond the Flush}

\author[1]{Pepe Vila}
\author[2]{Andreas Abel}
\author[1]{Marco Guarnieri}
\author[3]{Boris K\"opf}
\author[2]{Jan Reineke}
\affil[1]{IMDEA Software Institute}
\affil[2]{Saarland University}
\affil[3]{Microsoft Research}

\maketitle

\begin{abstract}
Flushing the cache, using instructions like \texttt{clflush} and \texttt{wbinvd}, is commonly proposed as a countermeasure against access-based cache attacks.
In this report, we show that several Intel caches, specifically the L1 caches in some pre-Skylake processors and the L2 caches in some post-Broadwell processors, leak information even after being flushed through \texttt{clflush} and \texttt{wbinvd} instructions.
That is, security-critical assumptions about the behavior of  \texttt{clflush} and \texttt{wbinvd}  instructions are incorrect, and countermeasures that rely on them should be revised.
\end{abstract}

\section{Introduction}\label{sec:intro} 

Caches are small, fast memories that bridge the latency gap between the CPU and the main memory.
Caches are critical for performance: they speed up computation by storing recently accessed data and reducing the interaction with main memory~\cite{memorywall1995}.

Caches are also critical from a security perspective since they are often shared (both temporally and spatially) across security domains.
This has inspired a multitude of covert~\cite{MauriceHello2017,lruleak2019} and side-channel attacks~\cite{Bernstein2004,Osvik2006,Tromer2010,YaromFlushReload2014,flushflush2016} in many different settings: web pages~\cite{OrenSpy2015,CachePortableGenkin2018,Vila19evictionsets}, OS processes~\cite{Hund2013,GrussTemplate2015}, mobile phones~\cite{LippARM2016,AutoLock2017}, virtual machines~\cite{LiuLLC2015,Irazoqui2015,MauriceHello2017}, and trusted environments~\cite{trustspy2016,cachezoom2017,sgxcache2017,sgxgrand2017,sgxmalware2017}.
Caches also have a prominent role in recent transient-execution~\cite{spectre18,meltdown2018,foreshadow2018,ridl2019,fallout2019,zombieload2019} attacks, which use caches as high-bandwidth exfiltration channels.

Classic cache-based attacks, such as Prime+Probe~\cite{Osvik2006} and Flush+Reload~\cite{YaromFlushReload2014}, leak information through the content of the cache, i.e., which memory blocks are accessed and stored in the cache.
Researchers, however, also explored leaks through the so-called cache's {\em control state}~\cite{cachequery2020}, i.e., the cache's replacement policy metadata.
Doychev et al.~\cite{cacheaudit2013} show that the lookup table preloading countermeasure used in some AES implementations may leak information through the pseudo-LRU (Least Recently Used) policy's state.
More recently, new attacks based on the cache's control state have been devised for high-bandwidth covert channels~\cite{lruleak2019}, bypassing Intel CAT's~\cite{intelrdt} isolation~\cite{dawg2018}, and performing practical side-channel attacks in modern L3 caches~\cite{samira2019}.

To protect a victim program against access-based cache attackers (i.e., attackers that can interact with the shared cache before and after the victim's execution), it suffices to prevent that the effects of the victim's execution on the cache cross the security domain boundary.
Given the belief that cache flushing (using the \texttt{clflush} and \texttt{wbinvd} instructions) cleanses the cache from any history-dependent information, cache flushing on context switches has been thought to provide---at least on single-threaded cores---resistance against access-based attacks. As an example, several works~\cite{
cryptoeprint:2019:613,
cacheinthecloud2013,
shielding2018,
duppel2013,
braun2015robust,
real2017spatial,
bazm2017side,
varadarajan2014scheduler,
aciiccmez2010new,
wistoff2020prevention
}
propose using cache flushing (sometimes in combination with other countermeasures and optimizations) to prevent cache covert and side-channel attacks.

Cache flushing has also been used as part of mitigations against recent microarchitectural vulnerabilities like L1TF~\cite{foreshadow2018} or CacheOut~\cite{cacheout2020}.
For instance, Intel suggests that cache flushing (using \texttt{clflush} and \texttt{wbinvd} instructions) and finer-grained flushing instructions like \texttt{IA32\_FLUSH\_CMD} (supported by new microcode updates when the \verb!L1D_FLUSH! processor flag is set) could be used to remove secrets from the L1D cache~\cite{intelforeshadow}.
Furthermore, processors for which the \verb!L1D_FLUSH! flag is enabled and which are affected by L1TF will automatically flush the L1D cache when executing the \verb!RSM! instruction that exits System Management Mode (SMM). Similar mechanisms have been recommended for protecting virtual machines and SGX enclaves on context switches~\cite{foreshadow2018, intelforeshadow}.

Previous work by Ge et al.~\cite{yourprocessorleaks2016,timeprotection19} investigates the effectiveness of flushing operations and shows that information persists in some microarchitectural components (specifically in the instruction cache, branch target buffer, branch history table, and translation lookahead buffer) even after flushing.
They also observe some {\em residual} leakage after flushing the data cache, and they associate it to the effects of data prefetchers.\looseness=-1

In work concurrent to ours, Wistoff et al.~\cite{wistoff2020prevention} show that, on the Ariane RISC-V core~\cite{ariane2019}, software solutions based on priming\footnote{RISC-V cache management still lacks standard flushing mechanisms.} (i.e., completely fill the cache with new data) are insufficient to mitigate cache covert channels. This is explained by the pseudo-random cache replacement policy implemented in Ariane.

We complement these results and demonstrate that cache flushing does \textit{not} cleanse the cache from history-dependent information in several Intel processors.
Even though cache flushing effectively eliminates leaks through the content of the cache (i.e., {\em which} memory blocks are stored in it), it does \textit{not} prevent leaks through metadata, specifically through the state of the cache replacement policy (i.e., {\em how} memory blocks are accessed).

Our results imply that flush instructions are insufficient to fully defeat access-based cache attacks in some Intel CPUs.

\section{Cache Control States Survive Flushing}\label{sec:nondeterminism}

Prior work~\cite{cachequery2020,nanobench2019} independently reports on nondeterministic behaviors, on several Intel processors, after flushing  caches with \texttt{wbinvd} and \texttt{clflush} instructions.
Specifically, nondeterministic behavior after flushing has been observed in the L1 caches of Sandy Bridge, Ivy Bridge, Haswell, and Broadwell CPUs,
and in the L2 caches of Skylake, Kaby Lake, and Coffee Lake CPUs.

Our hypothesis is that the nondeterminism is due to the persistent control state of the cache replacement policy. Specifically:
\begin{asparaitem}
\item  {\em the control state is not modified after a flush operation} whereas the lines in the cache are invalidated;
\item  {\em insertion of new blocks in the cache does not fully override the control state whenever there are invalid lines}.
\end{asparaitem}
If our hypothesis holds, then the common assumption that instructions like \texttt{clflush} and \texttt{wbinvd} cleanse all information from the cache is incorrect.
This would also indicate that using flush operations to prevent access-based attacks is insufficient.

Our approach to validate the aforementioned hypothesis is the following:
(1)~First, we fill the cache set with associativity many known memory blocks $I_0 \dots I_{n-1}$; we call the resulting state $s_0$.
(2)~Then, we perform additional cache hits to bring the cache control state into an arbitrary known state~$s_i$.
(3)~Finally, we invalidate the cache contents (by executing \texttt{wbinvd} or several \texttt{clflush} instructions) and refill the cache with different memory blocks $I'_0 \dots I'_{n-1}$; we call the resulting state~$s'_i$.
If our hypothesis is {\em true}, and the control state withstands flush operations, the control state $s'_i$ will depend on the previous control state $s_i$.

\section{Validating the Hypothesis}
In this section, we empirically evaluate our hypothesis that information from the cache's control state survives the flush operations on 11 different Intel processors from different generations.

We start by introducing the tools and setup of our evaluation. We continue by discussing two in-depth examples targeting the L1 PLRU cache from an i7-4790 CPU, and the L2 Quad-age LRU cache from an i5-6500. We conclude our evaluation by presenting a summary of all our findings.

\subsection{Tools and Setup}
In our evaluation, we use two tools to interact with caches: \emph{CacheQuery}~\cite{cachequery2020} and the \emph{nanoBench Cache Analyzer}~\cite{nanobench2019}.
Both tools provide a clean interface and low-noise environment for probing caches, liberating the user from dealing with intricate details such as the virtual-to-physical memory mapping, cache slicing, set indexing, cache filtering, and other sources of interferences or measurement noise, thus enabling a ``civilized'' interaction with an individual cache set.

Namely, users can specify a cache set (say: set 63 in the L2 cache) and a pattern of memory accesses (say: $I_0 \cdot I_1 \cdot I_2 \cdot I_0 \cdot I_1 \cdot I_2$), and they receive as output a sequence (say: \texttt{Miss Miss Miss Hit Hit Hit}) representing the hits and misses produced by a sequence of memory loads to addresses that are mapped into the specified cache set and that follow the specified pattern.

\paragraph{Example}
For instance, we can identify the Least Recently Used (LRU) block of a cache containing blocks $I_0 ~ \dots ~ I_7$ as follows:
we cause an eviction by accessing $I_8$ (a block not in the cache) and then check whether accessing block $I_j$, where  $0 \leq j \leq 7$, produces a hit or a miss.
Note that one would need to reset the cache state each time and access $I_8 \cdot I_j$ for all $0 \leq j \leq 7$ to determine which block has been evicted.
If accessing $I_8 \cdot I_1$ causes a cache miss for $I_1$, this shows that $I_1$ has been evicted by $I_8$ and thus must have been the LRU block prior to the access to $I_8$.

\begin{figure}[h]
	\centering
	\footnotesize
	\begin{forest}
	rounded/.style={circle, draw},
	squared/.style={rectangle, minimum width=7mm, minimum height=7mm, draw, anchor=south}
	[{0}, for tree=rounded
		[{0}, if n=1{edge=->}{}
			[{1}, if n=1{edge=->}{}
				[{$I_0$}, squared, if n=2{edge=->}{}]
				[{$I_1$}, squared, if n=2{edge=->}{}, fill=lightgray]
			]
			[{1}, if n=1{edge=->}{}
				[{$I_2$}, squared, if n=2{edge=->}{}]
				[{$I_3$}, squared, if n=2{edge=->}{}]
			]
		]
		[{0}, if n=1{edge=->}{}
			[{1}, if n=1{edge=->}{}
				[{$I_4$}, squared, if n=2{edge=->}{}]
				[{$I_5$}, squared, if n=2{edge=->}{}]
			]
			[{1}, if n=1{edge=->}{}
				[{$I_6$}, squared, if n=2{edge=->}{}]
				[{$I_7$}, squared, if n=2{edge=->}{}]
			]
		]
	]
	\end{forest}
	\caption{Cache state $s_1$ of an 8-way PLRU cache set after filling the cache with blocks $I_0~\dots~I_7$ and then accessing the reset sequence $I_0 \cdot I_2 \cdot I_4 \cdot I_6$. Arrows point to the LRU block, in this case $I_1$.}
\label{fig:plru-initial}
\end{figure}
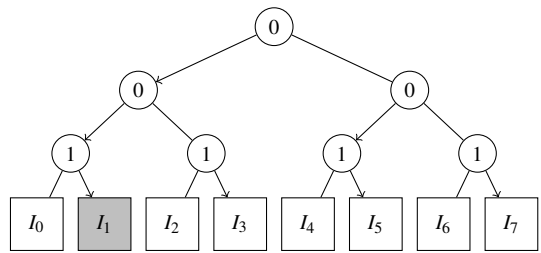

\subsection{Example 1: L1's Tree-based PLRU}\label{sec:evaluation:l1-plru}

As confirmed by prior research~\cite{cachequery2020,nanobench2019}, many Intel CPUs implement a tree-based Pseudo-LRU replacement policy in their L1 caches.

\paragraph{Replacement Policy}
Tree-based PLRU is a well known approximation of the LRU policy.
In an $n-$way cache, PLRU maintains $n-1$ control bits for each cache set.
Conceptually, a PLRU cache set can be seen as a balanced binary tree, in which the control bits correspond to the internal nodes, and the leaves correspond to the $n$ cache lines of the respective cache set.
A control bit valued 0 represents an arrow pointing to the left child, a 1 represents an arrow pointing to the right child. See Figure~\ref{fig:plru-initial} for an example of a PLRU tree state---with $I_0 \dots I_7$ initially in cache lines $0$ to $7$---after accessing the {\em reset sequence} $I_0 \cdot I_2 \cdot I_4 \cdot I_6$, which moves the cache from {\em any} state into a fixed known state~\cite{kisielewicz2015,kudlacik2012}.

Upon a cache miss, the block to replace is identified by following the arrows from the root node.
Upon a cache hit, all the ancestors of the accessed cache line update their arrows to point towards the opposite direction of the leaf, thereby protecting the accessed cache line from eviction in the near future.
When a new block is inserted, the arrows are updated as in the hit operation.

As a tree with $n$ leafs has $n-1$ internal nodes, corresponding to the PLRU bits in our example, the total number of control states for PLRU is $2^{n-1}$, where $n$ is the associativity or the number of ways of the cache. Thus, for an 8-way PLRU cache we have $128$ different control states.

\paragraph{Experiment}

We now proceed to test our hypothesis, as described in Section~\ref{sec:nondeterminism}, for the L1 cache of an i7-4790 processor.

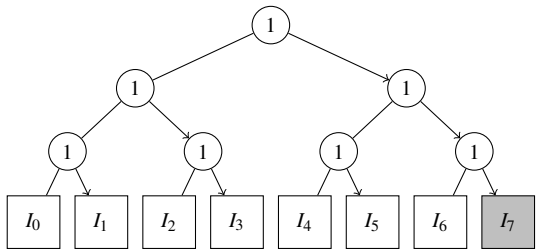
\begin{figure}[h]
	\centering
	\footnotesize
	\begin{forest}
    rounded/.style={circle, draw},
	squared/.style={rectangle, minimum width=7mm, minimum height=7mm, draw, anchor=south}
	[{1}, for tree=rounded
		[{1}, if n=2{edge=->}{}
			[{1}, if n=2{edge=->}{}
				[{$I_0$}, squared, if n=2{edge=->}{}]
				[{$I_1$}, squared, if n=2{edge=->}{}]
			]
			[{1}, if n=2{edge=->}{}
				[{$I_2$}, squared, if n=2{edge=->}{}]
				[{$I_3$}, squared, if n=2{edge=->}{}]
			]
		]
		[{1}, if n=2{edge=->}{}
			[{1}, if n=2{edge=->}{}
				[{$I_4$}, squared, if n=2{edge=->}{}]
				[{$I_5$}, squared, if n=2{edge=->}{}]
			]
			[{1}, if n=2{edge=->}{}
				[{$I_6$}, squared, if n=2{edge=->}{}]
				[{$I_7$}, squared, if n=2{edge=->}{}, fill=lightgray]
			]
		]
    ]
\end{forest}
\caption{Cache state $s_2$ of PLRU cache after marking $I_7$ as the LRU block, by accessing sequence $I_0 \cdot I_2 \cdot I_4 \cdot I_6 \cdot I_4 \cdot I_0$.}
\label{fig:plru-marklru}
\end{figure}

First, we fill the cache with blocks $I_0 ~ \dots ~ I_7$, which on an empty cache are inserted from left to right, and we bring the cache's control state to any desired state, for example, to state~$s_2$ by further accessing the sequence $I_0 \cdot I_2 \cdot I_4 \cdot I_6 \cdot I_4 \cdot I_0$ (see Figure~\ref{fig:plru-marklru}).

Then, we continue by invalidating the cache with a \verb!wbvind! instruction, and refilling the cache with associativity many new memory blocks $I_8 \dots I_{15}$, which again are inserted from left to right.

\begin{figure}[h]
	\centering
	\footnotesize
	\begin{forest}
    rounded/.style={circle, draw},
	squared/.style={rectangle, minimum width=7mm, minimum height=7mm, draw, anchor=south}
	[{1}, for tree=rounded
		[{1}, if n=2{edge=->}{}
			[{1}, if n=2{edge=->}{}
				[{$I_8$}, squared, if n=2{edge=->}{}]
				[{$I_9$}, squared, if n=2{edge=->}{}]
			]
			[{1}, if n=2{edge=->}{}
				[{$I_{10}$}, squared, if n=2{edge=->}{}]
				[{$I_{11}$}, squared, if n=2{edge=->}{}]
			]
		]
		[{1}, if n=2{edge=->}{}
			[{1}, if n=2{edge=->}{}
				[{$I_{12}$}, squared, if n=2{edge=->}{}]
				[{$I_{13}$}, squared, if n=2{edge=->}{}]
			]
			[{1}, if n=2{edge=->}{}
				[{$I_{14}$}, squared, if n=2{edge=->}{}]
				[{$I_{15}$}, squared, if n=2{edge=->}{}, fill=lightgray]
			]
		]
    ]
	\end{forest}
	\caption{Resulting cache state $s'_2$ after invalidating state $s_2$ and refilling the cache with new blocks $I_8 \dots I_{15}$. On an empty cache blocks are inserted from left to right.}
\label{fig:plru-refill}
\end{figure}
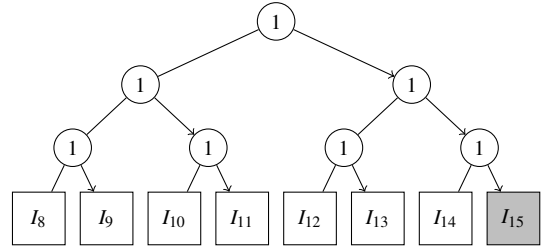

At this point, we validate our hypothesis by comparing the resulting control state $s'_2$ (see Figure~\ref{fig:plru-refill}) with $s_2$ (see Figure~\ref{fig:plru-marklru}).

For that, we rely on CacheQuery to probe the cache state and we verify that the eviction order of $s'_2$'s blocks is $I_{15}, I_{11}, I_{13}, I_9, I_{14}, I_{10}, I_{12}, I_{8}$, which uniquely corresponds to a control state identical to that of $s_2$.

After repeating the experiment for all possible states, we conclude that our hypothesis holds and that the control state survives both the flushing and the insertion of new blocks.

We observe the same effect when replacing the \verb!wbinvd! instruction with a sequence of  \verb!clflush! instructions for each block $I_0 \dots I_7$, or when writing \verb!1! to \verb!IA32_FLUSH_CMD! MSR.

\subsection{Example 2: L2's Quad-age LRU}\label{sec:evaluation:l2-quad-age-lru}

Modern Intel CPUs have an L2 4-way cache with an undocumented replacement policy that has only recently been reverse engineered~\cite{cachequery2020,nanobench2019}.

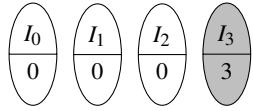
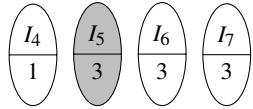
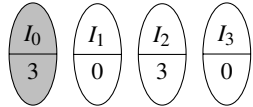
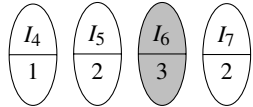
\begin{figure}
\begin{subfigure}{\columnwidth}
	\centering
	\small
	\begin{forest}
	squared/.style={ellipse split, minimum width=5mm, minimum height=10mm, draw, anchor=south}
	[{}, for descendants={no edge, squared}
		[{$I_0$ \nodepart{lower} 0}]
		[{$I_1$ \nodepart{lower} 0}]
		[{$I_2$ \nodepart{lower} 0}]
		[{$I_3$ \nodepart{lower} 3}, fill=lightgray]
	]
	\end{forest}
	\caption{Cache state $s_1$ on a 4-way {\em New1} cache after filling an empty (or invalidated) cache set with blocks $I_3 \dots I_0$ and further accessing sequence $I_0 \cdot I_1 \cdot I_2 \cdot I_0 \cdot I_1$.}
	\label{fig:quadlru-s1}
	\end{subfigure}
\begin{subfigure}{\columnwidth}
	\centering
	\small
	\begin{forest}
	squared/.style={ellipse split, minimum width=5mm, minimum height=10mm, draw, anchor=south}
	[{}, for descendants={no edge, squared}
		[{$I_4$ \nodepart{lower} 1}]
		[{$I_5$ \nodepart{lower} 3}, fill=lightgray]
		[{$I_6$ \nodepart{lower} 3}]
		[{$I_7$ \nodepart{lower} 3}]
	]
	\end{forest}
	\caption{Resulting cache state $s'_1$ after invalidating state $s_1$ and refilling the cache set with new blocks $I_7 \dots I_4$. On an empty cache blocks are inserted from right to left.}
	\label{fig:quadlru-s1-unknown}
	\end{subfigure}
\begin{subfigure}{\columnwidth}
	\centering
	\small
	\begin{forest}
	squared/.style={ellipse split, minimum width=5mm, minimum height=10mm, draw, anchor=south}
	[{}, for descendants={no edge, squared}
		[{$I_0$ \nodepart{lower} 3}, fill=lightgray]
		[{$I_1$ \nodepart{lower} 0}]
		[{$I_2$ \nodepart{lower} 3}]
		[{$I_3$ \nodepart{lower} 0}]
	]
	\end{forest}
	\caption{Cache state $s_2$ on a 4-way {\em New1} cache set after filling an empty (or invalidated) cache set with blocks $I_3 \dots I_0$ and further accessing sequence $I_0 \cdot I_1 \cdot I_2 \cdot I_0 \cdot I_1 \cdot I_3 \cdot I_1$.}
	\label{fig:quadlru-s2}
	\end{subfigure}
\begin{subfigure}{\columnwidth}
	\centering
	\small
	\begin{forest}
	squared/.style={ellipse split, minimum width=5mm, minimum height=10mm, draw, anchor=south}
	[{}, for descendants={no edge, squared}
		[{$I_4$ \nodepart{lower} 1}]
		[{$I_5$ \nodepart{lower} 2}]
		[{$I_6$ \nodepart{lower} 3}, fill=lightgray]
		[{$I_7$ \nodepart{lower} 2}]
	]
	\end{forest}
	\caption{Resulting cache state $s'_2$ after invalidating state $s_2$ and refilling the cache set with new blocks $I_7 \dots I_4$. On an empty cache blocks are inserted from right to left.}
	\label{fig:quadlru-s2-unknown}
	\end{subfigure}
\caption{List of states for L2 QLRU variant's experiment.}
\end{figure}

\paragraph{Replacement Policy}
This policy is an instance of Quad-Age LRU~\cite{intelhotchips2012}, called {\em New1} in~\cite{cachequery2020} and QLRU\_H00\_M1\_R2\_U1 in~\cite{nanobench2019}.
The policy keeps two bits of control state (or metadata) per line, which can be interpreted as associating one of four possible ages with each line (0 to 3), hence the name. See Figure~\ref{fig:quadlru-s1} for an example.\looseness=-1

Upon a cache hit, the age of the accessed block is set to~$0$.
Upon a cache miss, the first line---starting from the left---with age $3$ is replaced, and the new block is inserted in this line with an initial age of $1$. Observe that empty (or invalidated) lines are filled from right to left, before considering blocks with age $3$.
After each memory access, the ages of the lines are normalized to ensure the presence of an age-3 cache line:
As long as there is no age-$3$ cache line, the ages of cache lines, except for the updated (or inserted) one, are incremented by~$1$. 

After normalization, any valid state has at least one age-3 cache line; and after a hit or miss, at least one cache line with age $0$ or $1$.
Thus, for a 4-way cache, we can count the total number of valid states---on a filled set---as $4^4 - 3^4 - 2^4 + 1 = 160$, where: $4^4$ is the state space size, $3^4$ are the states without age-3 lines, $2^4$ are the states without ages $1$ and $0$, and $1$ is the control state $\{2,2,2,2\}$ that we subtracted twice.

\paragraph{Experiment}

To validate our hypothesis that the control state survives cache-flushing operations, we perform an experiment, similar to the one described in Section~\ref{sec:evaluation:l1-plru}, for the L2 cache of a Core i5-6500 processor.

First, we fill the cache with blocks $I_3 \dots I_0$, which on an empty cache are inserted from right to left, and bring the cache's control state to a specific state, for example, to $s_1$ by further accessing the sequence $I_0 \cdot I_1 \cdot I_2 \cdot I_0 \cdot I_1$ (see Figure~\ref{fig:quadlru-s1}).

Then, we continue by invalidating the cache with a \verb!wbinvd! instruction, and by refilling the cache with associativity many new memory blocks $I_7 \dots I_4$, which again are inserted from right to left.

Unfortunately, in this case, comparing $s'_1$ (see Figure~\ref{fig:quadlru-s1-unknown}) and $s_1$ (see Figure~\ref{fig:quadlru-s1}) is not enough for validating our hypothesis, because the insertion of new blocks modifies the control state.
We later provide a precise description of this behavior.

Instead, we check whether two different initial control states~$s_1$ and $s_2$ (see Figure~\ref{fig:quadlru-s2}) result in two different control states~$s'_1$ and $s'_2$ (see Figure~\ref{fig:quadlru-s2-unknown}), causing different eviction orders.
Namely, we use CacheQuery to test that $s'_1$ first evicts~$I_5$ and $s'_2$ first evicts $I_6$.

After repeating the experiment for all possible states, we conclude that our hypothesis holds and that the control state \emph{partially} survives the flushing and the insertion of new blocks, i.e., some initial control states can be distinguished after flushing and inserting new blocks, while other can not.

Note that we observe the same effect\footnote{Invalidation with \texttt{IA32\_FLUSH\_CMD} is available only for L1 caches.} when replacing the \verb!wbinvd! instruction with a sequence of  \verb!clflush! instructions for each block $I_0 \dots I_3$.

\paragraph{Reverse Engineering the Insertion Logic}
In order to reverse engineer the insertion logic when invalid blocks are present, we obtain all the resulting control states, after invalidating and refilling the cache, from the $160$ possible initial control states.

For this, we set the cache set into a given control state, invalidate the cache set, insert associativity many new blocks, and probe the cache until we uniquely identify its resulting control state. Note that since the probing is destructive, we often require to redo all these steps.

The probing works as follows: (1)~Start with the complete set of states; (2)~Perform a random memory access and eliminate all the states that are inconsistent with the observation (i.e., hit or miss) according to the replacement policy; (3)~Repeat step 2 until we are left with a single possible state.

Once we obtained the mapping from the $160$ {\em pre-flush} control states to the {\em post-refill} control states, we manually inferred the following rules that are fully consistent with the mapping:
\begin{itemize}
	\item Invalidation does not reset the cache control state;
\item A new block's age is set to $0$ if the invalidated age was $0$, and it is set to $1$ otherwise.
	\item Normalization occurs as described earlier, and does not discriminate between valid and invalid lines.
\end{itemize}

Interestingly, with the simple refilling we use (i.e., $I_7 \cdot~\dots~\cdot I_4$), the set of {\em leaked} states is reduced to a subset of only $11$ different control states.
We do not explore whether more complicated insertions---with interleaved hits---can increase the size of this subset, and, therefore, increase the leakage.

\subsection{Experimental Results}\label{sec:evaluation:results}

In this section we test how the \verb!wbinvd! and \verb!clflush! operations affect the cache's control state in all the cache levels of 11 different Intel processors from different generations.
For each of the processors and cache levels, we performed tests similar to those explained in Sections~\ref{sec:evaluation:l1-plru}--\ref{sec:evaluation:l2-quad-age-lru}.

We quantify how much information survives the flush operation. For this, let the random variable $S$ be a uniform distribution of initial control states, and the random variable $O$ be the control states after a flush operation. The mutual information $I(S;O)=H(S)-H(S|O)$ captures how many bits are leaked.

For L1's PLRU, $H(S)=\log_2 128=7$ and $H(S|O)=0$, given that we are able to uniquely identify all the initial states. Hence we find that the leakage is $7$ bits.

For L2's {\em New1 (QLRU\_H00\_M1\_R2\_U1)}, $H(S)=\log_2 160=7.32$ and $H(S|O)$ requires more fine grained information about the joint probability distribution, which we are able to obtain from the $160$ transitions. We compute a leakage of $3.17$ bits.

Observe also that if there is only one observation (i.e., the resulting state after a flush is always the same) for any initial state, then $H(S|O)=H(S)$ and therefore the mutual information $I(S;O)$ is $0$ indicating that there is no leakage.

\begin{table}[h!]
\centering
\scriptsize
\begin{tabular}{c c c c}
\toprule
\textbf{CPU} & \textbf{Cache level} & \textbf{Assoc.} & \textbf{Leakage (bits)} \\ \midrule[0.08em]
\multirow{3}{*}{
\shortstack[c]{
\textit{Core i5-750}\\
\textit{(Nehalem)}
}
} & L1 & $8$ & $0$ \\& L2 & $8$ & $0$ \\& L3 & $16$ & $0$ \\\midrule[0.01em]
\multirow{3}{*}{
\shortstack[c]{
\textit{Core i5-650}\\
\textit{(Westmere)}
}
} & L1 & $8$ & $0$ \\& L2 & $8$ & $0$ \\& L3 & $16$ & $0$ \\\midrule[0.01em]
\midrule[0.01em]
\multirow{3}{*}{
\shortstack[c]{
\textit{Core i7-2600}\\
\textit{(Sandy Bridge)}
}
} & L1 & $8$ & {\color{red} $7$} \\& L2 & $8$ & $0$ \\& L3 & $16$ & $0$ \\\midrule[0.01em]
\multirow{3}{*}{
\shortstack[c]{
\textit{Core i5-3470}\\
\textit{(Ivy Bridge)}
}
} & L1 & $8$ & {\color{red} $7$} \\& L2 & $8$ & $0$ \\& L3 & $12$ & $0$ \\\midrule[0.01em]
\multirow{3}{*}{
\shortstack[c]{
\textit{Core i7-4790}\\
\textit{(Haswell)}
}
} & L1 & $8$ & {\color{red} $7$} \\& L2 & $8$ & $0$ \\& L3 & $16$ & $0$ \\\midrule[0.01em]
\multirow{3}{*}{
\shortstack[c]{
\textit{Core i5-5200U}\\
\textit{(Broadwell)}
}
} & L1 & $8$ & {\color{red} $7$} \\& L2 & $8$ & $0$  \\& L3 & $12$ & $0$ \\\midrule[0.01em]
\midrule[0.01em]
\multirow{3}{*}{
\shortstack[c]{
\textit{Core i5-6500}\\
\textit{(Skylake)}
}
} & L1 & $8$ & $0$ \\& L2 & $4$ & {\color{orange} $3.17$} \\& L3 & $12$ & $0$ \\\midrule[0.01em]
\multirow{3}{*}{
\shortstack[c]{
\textit{Core i7-8550U}\\
\textit{(Kaby Lake)}
}
} & L1 & $8$ & $0$ \\& L2 & $4$ & {\color{orange} $3.17$} \\& L3 & $16$ & $0$ \\\midrule[0.01em]
\multirow{3}{*}{
\shortstack[c]{
\textit{Core i7-8700K}\\
\textit{(Coffee Lake)}
}
} & L1 & $8$ & $0$ \\& L2 & $4$ & {\color{orange} $3.17$} \\& L3 & $16$ & $0$ \\\midrule[0.01em]
\midrule[0.01em]
\multirow{3}{*}{
\shortstack[c]{
\textit{Core i3-8121U)}\\
\textit{(Cannon Lake)}
}
} & L1 & $8$ & $0$ \\& L2 & $4$ & $0$ \\& L3 & $16$ & $0$ \\\midrule[0.01em]
\multirow{3}{*}{
\shortstack[c]{
\textit{Core i5-1035G1)}\\
\textit{(Ice Lake)}
}
} & L1 & $12$ & $0$ \\& L2 & $8$ & $0$ \\& L3 & $12$ & $0$ \\\bottomrule
\end{tabular}
\caption{List of evaluated processors and cache levels. The leakage in bits correspond to the mutual information $I(S;O)$.
}
\label{table:results}
\end{table}

Table~\ref{table:results} reports all our findings, which we summarize here:
\begin{asparaitem}
\item For pre-Skylake processors (except for Nehalem and Westmere), the control state of L1 caches persists even after flushing operations (\verb!wbinvd!, sequences of  \verb!clflush!, and writing \verb!1! to the \verb!IA32_FLUSH_CMD! MSR).
\item For processors between Skylake and Coffee Lake,  the control state of L2 caches persists even after flushing operations (through \verb!wbinvd! and sequences of  \verb!clflush!).
\item According to our experiments, it seems that the control state is wiped out on flushing for the most recent CPU families, like Cannon Lake and Ice Lake.

\end{asparaitem}

\section{Discussion}
In this section, we briefly discuss possible implications of, and possible solutions to, \attack{}.

\paragraph{Access-based Attacks}
Access-based cache attackers\footnote{We dismiss more powerful trace-based attackers, since general security guidelines already recommend to disable hyper-threading for sensitive computations.} monitor their own cache activity to infer activity from a victim, namely, which cache lines or cache sets the victim accessed. While this provides a powerful primitive, it only exploits one dimension: {\em what} data is accessed.
Knowledge about the control state provides access to a new dimension: {\em how} data is accessed.

While observing the control state is difficult, it is possible for attackers with low-level control of the system~\cite{cachezoom2017,sgxstep2017}, and for more realistic user-space attackers~\cite{lruleak2019,dawg2018,samira2019}.

However, these attacks were only possible for an adversary that shared memory, and hence the cache lines and control state, with the victim.
\attack{} breaks this assumption and enables an access-based attacker to leak the control state of the cache in non-shared memory scenarios. This is possible by flushing and refilling the cache with the attacker's own contents, before probing the state.

\paragraph{Extending Prime+Probe}
Here, we briefly describe how to extend an L1 Prime+Probe attack, which can leak the cache sets accessed by a victim's computation, with \attack{}  to work even when the complete L1 is invalidated (e.g. via \texttt{IA32\_FLUSH\_CMD}) after each victim's execution.
First, the attacker primes all the cache sets of interest and accesses a reset sequence on each of them to set a known control state (say: Figure~\ref{fig:plru-marklru}'s $s_2$).
Next, the victim executes the secret computation and, before terminating, invalidates all the cache sets.
Finally, the attacker probes each cache set by (1) filling it with controlled data (say: $I_0 \dots I_7$, which as seen does not modify the control state), (2) causing an eviction, and (3) measuring whether $I_7$ (according to our $s_2$ example) is still in the cache.

For a given cache set, after priming, any victim access will necessarily cause at least one eviction thereby changing the control state by flipping away all the arrows from $I_7$'s previous slot.
Hence, when the attacker probes the cache, this will result in evicting a block different from $I_7$, whose access will result in a cache hit.

Observe that while multiple victim accesses might bring the cache set into the initial control state, thereby resulting in potential false positives, this event is unlikely. Similarly, if the cache is instead invalidated {\em before} the victim's computation, the misses caused by the initial insertions would not modify the control state, but any subsequent misses and hits would.

\paragraph{Distinguishing Sequences}
The enhanced Prime+Probe example, as well as prior attacks~\cite{lruleak2019,samira2019}, exploit a 1-bit leak to distinguish between 2 control states based on whether a specific access causes a hit or a miss.
Here, we propose a generalization by using the notion of {\em distinguishing sequences} from automata theory~\cite{soucha2014}.

For a specific cache replacement policy, the output (i.e., the observed sequence of hits and misses) produced by a sequence of memory accesses induces a partition $P$ over the set of all initial control states, where two control states are in the same set whenever they result in the same output.
Given the automata model of the policy, which can be obtained for instance from~\cite{cachequery2020}, we can compute a distinguishing sequence $\overline{d}$ that produces the finest possible partition $P$.
This means that, using $\overline{d}$, one can distinguish up to $|P|$ subsets of initial control states in a single experiment.\footnote{This quantity is closely related to the so called cache extraction~\cite{canones2017}.}
Distinguishing sequences capture the destructive nature of cache probing and enable the computation of optimal strategies. A set can not be further refined after the probing has destroyed all the initial information.

\begin{figure}[h]
	\centering
\forestset{
		my edge label/.style 2 args={
			edge label={node[midway, font=\sffamily\scriptsize, #1]{#2}},
		},
	}
	\begin{forest}
	[{$S$},
		[{$S$},my edge label={right}{$I_4/M$}
		[{$\{s_0,s_5\}$}, my edge label={above}{$I_0/M$}
			[{$\{s_0,s_5\}$}, my edge label={left}{$I_1/*$}
				[{$\{s_0\}$}, my edge label={left}{$I_2/M$}]
				[{$\{s_5\}$}, my edge label={right}{$I_2/H$}]
			]
		]
		[{$\{s_1,s_2,s_3,s_4,s_6,s_7\}$}, my edge label={above}{$I_0/H$}
			[{$\{s_2,s_4\}$}, my edge label={left}{$I_1/M$}
				[{$\{s_4\}$}, my edge label={left}{$I_2/M$}]
				[{$\{s_2\}$}, my edge label={right}{$I_2/H$}]
			]
			[{$\{s_1,s_3,s_6,s_7\}$}, my edge label={right}{$I_1/H$}
				[{$\{s_1,s_6\}$}, my edge label={left}{$I_2/M$}]
				[{$\{s_3,s_7\}$}, my edge label={right}{$I_2/H$}]
			]
		]
		]]
	\end{forest}
	\caption{Example of an optimal preset distinguishing sequence for a 4-way PLRU cache. Nodes are labelled with the set of control states consistent with the observations. Edges are labelled with pairs $I/O$ where $I$ is the accessed memory block and $O \in \{H,M\}$ is the corresponding observation, where $H$ represents a cache hit and $M$ a cache miss.}
\label{fig:distinguish}
\end{figure}
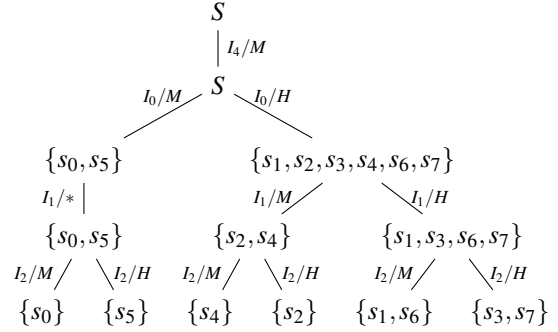

We are able to compute both {\em preset}---i.e., non-adaptive---and adaptive sequences, but for simplicity we only illustrate the preset example\footnote{While for PLRU both approaches lead to equal $|P|$, this is not necessarily true for other policies.}.
Consider, for instance, a 4-way PLRU cache with initial content $I_0 \dots I_3$ and initial control state $s_i \in S$, in this example $S=\{s_0,\dots,s_7\}$.
The distinguishing sequence $\overline{d} = I_4~\cdot~I_0~\cdot~I_1~\cdot~I_2$ produces an optimal partition $P$ consisting of $6$ subsets. Figure~\ref{fig:distinguish} shows how the resulting hit/miss pattern determines to which subset of $S$ the initial control state $s_i$ belongs to.

\paragraph{Cache Partitioning}

Cache partitioning~\cite{pageplacement1992} was initially proposed to improve predictability by reducing cache contention.
Since then, several works~\cite{cachecoloring2011,stealthmem2012,catalyst2016,timeprotection19} have proposed the use of cache partitioning to also mitigate cache leakage.
For instance, mechanisms like Intel's CAT~\cite{intelrdt} allow the partitioning of the cache's ways.
In contrast, mechanisms like page coloring split the cache sets among untrusted parties.
While Intel's CAT is known to leak through the cache replacement policy~\cite{dawg2018}, for page coloring the question remains open.
Thus, we like to point to modern adaptive policies, as described in \cite{ibpolicy2013,cachequery2020}, as promising candidates for answering in the affirmative the question on cross cache set information leakage.

\paragraph{Countermeasures}
Software mitigations would involve replacing (or extending) flush instructions on context switches with accesses to reset sequences---that bring the cache into a fix control state. Unfortunately, this requires knowledge of the specific cache contents at that point (to cause hits), or an even higher overhead due to the additional misses.
We refer to~\cite{ReinekeGBW07} for a detailed analysis---of LRU, FIFO, PLRU, and MRU replacement policies---providing tight bounds on the effort for establishing a desired cache set state.

\section{Conclusion}
We evaluate the behavior of cache flushing instructions on several Intel processors and conclude that they do not properly cleanse all the information stored in the cache. Specifically, we show that in some caches the control state survives, allowing information leakage beyond cache flushes.
We point out that countermeasures relying solely on flush instructions should be revised.

\section*{Disclosure}
We first discussed our observations about the cache control state surviving flush operations in November 2019. 

We reported our findings to Intel's PSIRT team on the 13th of April 2020, after having confirmed and understood the source of leakage.

Intel responded to us on the 19th of May 2020, concluding that the issue does not pose more risks than traditional cache side-channels, and thus recommending their best practice guidelines against side-channel attacks~\cite{intelguideline}.

 
\bibliographystyle{plain}

\end{document}